\documentclass[preprint,
               prd,nofootinbib,superscriptaddress,
               showpacs,preprintnumbers,
               amsmath,amssymb]{revtex4}
\begin{document}

\preprint{INR/TH-20-2005}

\title{
  Lorentz-violating brane worlds and cosmological perturbations
}

\author{M.V. Libanov and V.A. Rubakov}

\affiliation{
Institute for Nuclear Research of the Russian Academy of Science,\\
60th October Anniversary prospect 7a, Moscow 117312, Russia
}

%\date{}

\begin{abstract}
We consider an inflating brane-world setup in which 4-dimensional
Lorentz-invariance is violated at high 3-momentum scale $P_{LV} \gg H$,
where $H$ is the inflationary Hubble parameter. We study massless scalar
field in this background as a model for cosmological perturbations.
Towards the end of inflation, the spectrum has both the standard,
4-dimensional part due to a brane-localized mode, and exotic, bulk induced
contribution. The suppression of the latter is power-law only,
$(H/P_{LV})^\alpha$, provided that there exist bulk modes with energies
$\omega \ll H$. Contrary to  general expectations, the exponent $\alpha$
may be smaller than 2, and even smaller than 1, depending on details of
the bulk geometry. Furthermore, the overall amplitude of the bulk-induced
perturbations is enhanced as compared to the standard part, so the effects
due to Lorentz-violation may dominate over the standard mechanism even for
$P_{LV} \gg H$.
\end{abstract}


\maketitle
%%%%%%%%%%%%%%%%%%%%%%%%%%%%%%%%%%%%%%%%%%%%%%%%%%%%%%%%%%%%%%%%%%%%%%%%%%%%%%%%%%%%%%%%

%%
%%%%%%%%%%%%%%%%%%%%%%%%%%%%%%%%%%%%%%%%%%%%%%%%%%%%%%%%%%%%%%%%%%%%%%%%
%\begin{document}

\section{Introduction and summary}
\label{sec:intro}

In brane-world scenarios, it is not inconceivable that four-dimensional
Lorentz-invariance is violated in the bulk~\cite{L1,csaki,sergd,sibir}.
Models of this sort provide an interesting framework~\cite{mlvr} for
addressing the cosmological ``trans-Planckian'' problem~\cite{1,trans,2}
of whether or not possible Lorentz-violation at high 3-momenta may affect
the predictions for cosmological perturbations generated at inflation. A
conservative possibility is that 3-momentum scale of Lorentz-violation,
$P_{LV}$, is much higher than the Hubble parameter towards the end of
inflation, $H$. It has been argued~\cite{2} that for $P_{LV} \gg H$, the
standard inflationary predictions should remain almost intact, although
there are fairly exotic four-dimensional counterexamples~\cite{1,cex}
based, e.g., on Corley--Jacobson dispersion relation~\cite{cj}. Barring
exotica, there is still some debate~\cite{2,clinej} on whether the effects
of Lorentz-violation, and ``heavy physics'' in general, are at best of
order $(H/P_{LV})^2$, or weaker suppression is possible.

In our previous discussion~\cite{mlvr} of the effects due to
Lorentz-violation on cosmological perturbations in the brane-world
framework, we introduced an inflating version of the setup of
Ref.~\cite{sergd} with Lorentz-violating bulk. A drawback of this approach
is that the bulk geometry is completely {\it ad hoc}; neither the source
of Lorentz-violation nor mechanism of inflation are specified. An
advantage, however, is that the behavior of quantum fields, in particular,
their initial state, are well under control, so the calculation of the
spectrum is unambiguous. The particular setup used in Ref.~\cite{mlvr} was
not interesting at $P_{LV} \gg H$ --- the corrections to the standard
predictions were suppressed as $\mbox{exp} \left(- \mbox{const} \cdot
\frac{P_{LV}}{H} \right)$ --- so we concentrated there on strong effects
occurring in a less conservative case, $P_{LV} \buildrel < \over {_{\sim}}
H$. In this paper we modify our setup in such a way that the effects due
to Lorentz-violating bulk are only power-law suppressed at $P_{LV} \gg H$.
Needless to say, the above remarks concerning the entire approach apply to
this work as well.

Our main findings are somewhat unexpected. First, the suppression of the
bulk-induced contribution, which is generated on top of the standard
spectrum, strongly depends on the bulk geometry, and may be  weaker than
$(H/P_{LV})^2$ and even $H/P_{LV}$, in obvious contradiction to the claim
of Ref.~\cite{2}.  Second, this contribution to the spectrum is enhanced
by $\epsilon^{-3}$ where $\epsilon$ is another small free parameter
inherent in our model. Thus, for small enough $\epsilon$ the effects due
to Lorentz-violating bulk may compete with, and even dominate over the
standard four-dimensional mechanism in spite of the hierarchy $P_{LV} \gg
H$. There is a simple reason for the enhancement: Lorentz-violating bulk
modes exit the cosmological horizon earlier, and hence get frozen at
higher amplitudes, as compared to brane-localized ones. Relatively large
perturbations in the bulk are then partially transferred to the brane due
to subsequent (but still occurring at inflationary stage)
mixing between bulk and brane modes.

In our model, the spectrum of additional, bulk-induced perturbations is
flat for $H=\mbox{const}$ (in accord with the scaling argument of
Ref.~\cite{mlvr}) and almost flat for inflationary Hubble parameter slowly
varying in time. However, the amplitude and tilt are determined by the
expansion rate at earlier stages of inflation, as compared to the standard
theory. In the slow roll scenario this would mean that this part of
perturbations has less tilted spectrum. A potentially observable property
that the primordial perturbations are a sum of two Gaussian fields with
different tilts (and amplitudes) appears to be an interesting feature of
our model, and likely a whole class of models with Lorentz-violating bulk.

Our overall conclusion is that in Lorentz-violating brane-world models,
the properties of cosmological perturbations generated at inflation may
strongly depend on dynamics in the bulk, even if 3-momentum scale of
Lorentz-violation on the brane largely exceeds the inflationary Hubble
parameter. Needless to say, this dynamics may naturally be entirely
different for, e.g., inflaton and graviton modes, so that the standard
relations between the scalar and tensor perturbations may be completely
(or partially) destroyed.

\section{Generalities}

\subsection{Background geometry}

Let us consider $(4+1)$-dimensional model with the coordinates $(t,
x^i,y)$, $i=1,2,3$. We choose the background five-dimensional metric as
follows,
\begin{equation}
ds^2 =[\alpha^2(y) d t^2 - \beta^2(y) a^2(t) d{\bf x}^2] - \alpha^2 (y)
dy^2
\label{1a.1*}
\end{equation}
%
%\begin{equation}
%ds^2 =a^2(\eta)[\alpha^2(y) d\eta^2 - \beta^2(y) d{\bf x}^2] -
%\alpha^2 (y) dy^2
%\label{1.1*}
%\end{equation}
where the coordinate choice for $y$ is made for convenience. There is a
single brane at
\[
y=y_B
\]
The warp factors $\alpha (y)$ and $\beta (y)$ are continuous across the
brane, while their derivatives $\partial _y\alpha \equiv \alpha '$ and
$\partial _y\beta \equiv \beta '$ are not. The static background, $a(t)
=\mbox{const}$, is not four-dimensinally Lorentz-invariant; it is this
case that has been considered in Refs.~\cite{csaki,sergd}, where is has
been shown that, with appropriate choice of the warp factors,
four-dimensional Lorenz-invariance still holds for brane modes. An
equivalent form of the metric is
\begin{equation}
ds^2 =a^2(\eta)[\alpha^2(y) d\eta^2 - \beta^2(y) d{\bf x}^2] - \alpha^2
(y) dy^2
\label{1.1*}
\end{equation}
where the conformal time $\eta$ is related to time $t$ in the usual way,
\[
dt = a(\eta) d\eta
\]
We will use both forms of the metric in what follows.

We choose the warp factors in such a way that both $\alpha(y)$ and $\beta
(y)$ are $Z_2$-symmetric across the brane, monotonically decrease towards
large $y$ with
\begin{equation}
\beta^{\prime \prime} > 0
  \label{b''}
\end{equation}
and decay away from the brane,
\[
\alpha(y), \beta(y) \to 0 \; , \;\;\;\; \mbox{as} \; y\to \infty
\]
so that the integrals in (\ref{unity}) are finite. For what follows it
is convenient to rescale the coordinates $x^i$ and $t$ in such a way that
\begin{equation}
\int_{y_B}^\infty~dy~\alpha^2 \beta = \int_{y_B}^\infty~dy~\beta^3
\label{unity}
  \end{equation}
  Furthermore, we assume that  the ratio of warp factors tends to a small
  constant,
  \begin{equation}
\frac{\alpha(y)}{\beta(y)} \to \epsilon \; , \;\;\;\; \mbox{as} \; y\to
\infty
\label{1.1++}
\end{equation}
\[
\epsilon \ll 1
\]
The parameter $\epsilon$ is a free small parameter of our model. The
static case with $\epsilon=0$ was discussed in Ref.~\cite{sergd}. Here we
will be interested in inflating background,
\begin{equation}
a(t) = \mbox{exp}\left(\int~H(t) dt\right)
\label{infl}
\end{equation}
where $H(t)$ is a slowly varying function of time.

Finally, we assume that $\beta (y)$ decays sufficiently slowly as $y \to
\infty$, so that
\begin{equation}
\frac{\beta^\prime (y)}{\beta (y)}, \frac{\beta^{\prime \prime} (y)}{\beta
(y)} \to 0 \; , \;\;\;\; \mbox{as} \; y\to \infty
\label{assum}
 \end{equation}
The latter assumption is the key property that differs our background from
 that considered in Ref.~\cite{mlvr}; the peculiar features of bulk
perturbations that we alluded to in Introduction can be traced back
precisely to this property. In fact, the main results of this paper hold
for milder requirements on the behavior of $\beta(y)$ as $y \to \infty$:
it is sufficient to assume that $\left(\frac{\beta^\prime (y)}{\beta
(y)}\right)^2$ and $\frac{\beta^{\prime \prime} (y)}{\beta (y)}$ tend to
constant values which are sufficiently small compared to $H^2$. In what
follows we concentrate on the case (\ref{assum}) for definiteness.

To illustrate the dependence on the bulk geometry, we will take as a
concrete example
\begin{eqnarray}
\beta(y) &=& \frac{1}{y^\kappa} \nonumber \\
\alpha^2 (y) &=& \beta^2 (y)\left( \frac{3\kappa + 1}{3\kappa -1} \cdot
\frac{y_B^2}{y^2} + \epsilon^2\right)
\label{ab}
\end{eqnarray}
where
\begin{equation}
\kappa > 1/3
\label{kappa}
\end{equation}
The numerical factor in (\ref{ab}) is chosen in such a way that
Eq.~(\ref{unity}) is satisfied modulo unimportant
$\epsilon^2$-corrections. We will see that another small parameter of our
model, $y_B$, determines the 3-momentum $P_{LV}$ at which the effects of
Lorentz-violating bulk are substantial for brane modes.

It is worth stressing that our setup is completely {\it ad hoc}; we are
not aware of any realistic brane construction that would produce this
geometry. As mentioned in Introduction, this is the major disadvantage of
the entire approach.

\subsection{Brane and bulk modes}

In this paper we consider a real massless scalar field $\Phi$ in the
background (\ref{1a.1*}), meant to model perturbations of gravitational
and/or inflaton field. The action is
\[
S_\Phi = \frac{1}{2} \int~d^5X~ \sqrt{g} g^{AB} \partial_A \Phi
\partial_B \Phi
\]
It is consistent with $Z_2$ symmetry across the brane to impose the
Neumann boundary condition,
\[
\Phi^\prime (y=y_B) = 0
\]
Indeed, this boundary condition applies to appropriately defined
gravitational perturbations~\cite{RS}, and also to free scalar
field~\cite{BG} in the Randall--Sundrum background. It is convenient to
introduce the fields
\[
\phi (t,x^i,y) = \beta^{3/2}(y) \cdot \Phi (X^A)
\]
and
\[
\chi (\eta, x^i, y) = a(\eta) \beta^{3/2}(y) \cdot \Phi (X^A)
\]
which obey the field equations, in terms of three-dimensional Fourier
harmonics,
\begin{equation}
\frac{\partial^2 \phi}{\partial t^2} + 3H \frac{\partial \phi}{\partial t}
+ U(y) P^2 (t) \phi + L_y \phi =0
\label{eqphi1}
\end{equation}
and
\begin{equation}
\ddot{\chi} - \frac{\ddot{a}}{a} \chi + U(y) k^2 \chi + a^2(\eta) L_y \chi
= 0
\label{7}
\end{equation}
Hereafter dot denotes $d/ d\eta$, $k$ is time-independent conformal
3-momentum, while
\[
P(t) = \frac{k}{a(t)}
\]
is physical 3-momentum,
\begin{equation}
U(y) = \frac{\alpha^2(y)}{\beta^2(y)}
\label{U}
\end{equation}
and
\[
L_y = -\frac{\partial^2}{\partial y^2} + V(y)
\]
with
\[
V(y) = \frac{3}{2}\frac{\beta^{\prime \prime}}{\beta} +
\frac{3}{4}\frac{\beta^{\prime 2}}{\beta^2}
\]
The boundary conditions for $\phi$ and $\chi$ are
\begin{equation}
\left( \phi^\prime - \frac{3}{2} \frac{\beta^\prime}{\beta}
\phi\right)_{y=y_B} = \left( \chi^\prime - \frac{3}{2}
 \frac{\beta^\prime}{\beta} \chi\right)_{y=y_B} = 0
\label{print7*}
  \end{equation}
  The field $\chi$ is canonically normalized: its action is
 \[
  S_\chi = \int~d\eta~d^3 x~dy \left(\frac{1}{2} \dot{\chi}^2 + \dots
  \right)
  \]
  where dots denote terms independent of $\dot{\chi}$.

Let us first discuss bulk modes, assuming for the time being that the
background is static,
\[
H =0
\]
According to our assumptions (\ref{b''}) and (\ref{assum}), the
potential $V(y)$ is positive and vanishes as $y \to \infty$, so the
spectrum of the operator $L_y$ is continuous and starts from
zero\footnote{In the model of Ref.~\cite{mlvr}, the spectrum of the
operator analogous to $L_y$ started from a large positive value (in the
case of large Lorentz-violation scale). This difference is behind
different results obtained in the present paper and in Ref.~\cite{mlvr}.}.
Recalling Eq.~(\ref{1.1++}), we find that the dispersion relation for
these modes is
\begin{equation}
\omega^2 = \epsilon^2 P^2 + \lambda^2
\label{continuum}
\end{equation}
where $\lambda^2$ are eigenvalues of $L_y$, which can be arbitrarily
small.

Let us now consider brane modes, still in static background. At $P^2 =0$
there is a zero mode,
\[
\phi_0 (y) =  \beta^{3/2}
\]
which is normalizable, since the integrals in (\ref{unity}) are assumed
to be finite. It is worth noting that in terms of the original filed
$\Phi$ this mode is constant along extra dimension. Now, at finite but
small $P$, this mode gets lifted; the third term in Eq.~(\ref{eqphi1}) can
be treated as perturbation, and for the (real part of) energy one finds
\[
\omega^2 = P^2 \cdot \frac{\int_{y_B}^\infty~dy~U(y)
|\phi_0|^2}{\int_{y_B}^\infty~dy~|\phi_0|^2}
\]
Making use of Eqs.~(\ref{unity}) and (\ref{U}), one obtains the
Lorentz-invariant dispersion relation,
\[
\omega^2 = P^2
\]
Thus, the theory on the brane is (almost) Lorentz-invariant at small
3-momenta.

This is not the whole story, however. At small but finite $P$, the would
be zero mode is embedded in the continuum of bulk modes: its energy is
larger than the lowest energy $\omega =\epsilon P$ of continuum modes, see
Eq.~(\ref{continuum}). Therefore, the brane mode is {\it quasi}-localized
even at low 3-momenta, i.e., it has finite width against escape into extra
dimension. This effect was found in Ref.~\cite{sergd}. The
quasi-localization is due to mixing with bulk modes, which comes from the
third term in Eq.~(\ref{eqphi1}). Introducing the overlap between the
(normalized) zero mode $\phi _0$ and continuum modes $\phi _\lambda $,
\begin{equation}
  I_\lambda = \int~dy U(y) \phi_0^* (y) \phi_\lambda(y)
\label{Ilambda}
  \end{equation}
  we estimate the width of the quasilocalized state as
\begin{equation}
  \Gamma (P) \sim P^2 \cdot |I_{\lambda = P}|^2
\label{width}
  \end{equation}
  The whole picture of weak Lorentz-violation on the brane at low
  3-momenta and strong Lorentz-violation at high 3-momenta works provided
that the warp factors are chosen in such a way that
\begin{equation}
\lambda |I_\lambda|^2 \to 0 \; , \;\;\; \mbox{as} \;\;
\frac{\lambda}{P_{LV}} \to 0
\label{Ineq}
\end{equation}
and
\[
\lambda |I_\lambda|^2 \sim 1 \; , \;\;\; \mbox{at} \;\; \lambda \sim
P_{LV}
\]
It is this case that we consider in what follows. Then, the width of
the brane mode is small compared to its energy at low $P$, but it becomes
comparable to energy $\omega = P$ at $P \buildrel > \over {_{\sim}}
P_{LV}$. At $P>P_{LV}$ even quasi-localized brane mode ceases to exist, so
four-dimensional Lorentz-invariance is indeed completely destroyed.

Let us illustrate these features by making use of the concrete form
(\ref{ab}) of the warp factors. In this case one has\footnote{The field
equation in this case coincides, up to notations and field redefinition,
with the equation considered in Ref.~\cite{sergd}. The results in the end
of this section agree with Ref.~\cite{sergd} where they overlap.}
\begin{eqnarray}
   V(y) &=&  \frac{1}{y^2} \cdot \left( \frac{9}{4} \kappa^2 + \frac{3}{2}
   \kappa \right) \nonumber \\
U(y) &=& \frac{3\kappa +1}{3\kappa -1} \cdot \frac{y_B^2}{y^2} +
   \epsilon^2
\end{eqnarray}
The appropriately normalized eigenfunctions of $L_y$ are the zero mode
\begin{equation}
\phi_0 (y) = C(\kappa)\cdot\frac{y_B^{\frac{3\kappa
-1}{2}}}{y^{\frac{3\kappa}{2}}}
\label{zeromode}
\end{equation}
and bulk modes
\begin{equation}
\phi_\lambda (y) = \frac{1}{\sqrt{2}} \cdot \sqrt{\lambda y} \cdot
\frac{J_{\nu-1} (\lambda y_B) Y_\nu (\lambda y) - Y_{\nu-1} (\lambda y_B)
J_\nu (\lambda y)}{\sqrt{J_{\nu-1} (\lambda y_B)^2 + Y_{\nu -1} (\lambda
y_B)^2}}
\label{bulkmodes}
 \end{equation}
 where $J_\nu$ and $Y_\nu$ are the Bessel functions and
\[
 \nu = \frac{3\kappa +1}{2}
 \]
 Note that because of (\ref{kappa}), one has
 \[
 \nu > 1
 \]
 In Eq.~(\ref{zeromode}) and below $C(\kappa)$ denotes unimportant
 constants of order 1, which may be different in different formulas, but
never are equal to zero or infinity, provided that $\kappa$ obeys
(\ref{kappa}).

At $\lambda \ll y_B^{-1}$ the overlap integral (\ref{Ilambda}) is equal to
\[
I_\lambda = C(\kappa) y_B^{\nu -1} \cdot \lambda^{\nu - \frac{3}{2}}
\]
so that
\begin{equation}
\lambda I^2_\lambda \sim (y_B \lambda)^{2\nu -2}
\label{lIl}
\end{equation}
and the estimate (\ref{width}) for the width of the quasi-localized mode
reads
\[
\Gamma = C(\kappa) P \cdot (P y_B)^{2\nu -2}
\]
The direct calculation of the width by techniques well known from
quantum mechanics~\cite{Landau} confirms this estimate. At low 3-momenta
this width is small compared to the real part of energy, $\omega = P$,
although the suppression is mild at $\nu$ close to 1. On the other hand,
the width becomes large and the quasi-localized mode disappears at
\begin{equation}
P \buildrel > \over {_{\sim}} P_{LV} = \frac{1}{y_B}
\label{ybP}
\end{equation}
We see that in this particular setup, the position of the brane determines
the length scale of Lorentz-violation for brane-based observer.

\section{Generation of brane-localized perturbations}

\subsection{Evolution of modes: zeroth order}

Let us now consider the scalar field in inflating setup, Eq.~(\ref{infl}).
It is convenient to work with the field $\chi(\eta, y)$ and decompose it
in eigenfunctions of the operator $L_y$,
\[
\chi (\eta, y) = \psi_0 (\eta) \phi_0 (y) + \int_0^\infty~d\lambda
\psi_\lambda (\eta) \phi_\lambda (y)
\]
From Eq.~(\ref{7}) one obtains the system of equations
\begin{eqnarray}
\ddot{\psi}_0 - \frac{\ddot{a}}{a} \psi_0 + k^2 \psi_0 &=& - k^2
\int~d\lambda~\psi_\lambda I_\lambda
\label{psi_0}\\
\ddot{\psi}_\lambda - \frac{\ddot{a}}{a} \psi_\lambda + \epsilon^2 k^2
\psi_\lambda  + a^2 \lambda^2 \psi_\lambda &=& - k^2 I^{*}_\lambda \psi_0
- k^2 \int~d\lambda^\prime ~\psi_{\lambda^\prime} I_{\lambda,
\lambda^{\prime}}
\label{psi_lambda}
\end{eqnarray}
where
\begin{equation}
I_{\lambda \lambda^\prime} = \int~dy~\phi_\lambda^* [U(y) - \epsilon^2]
\phi_{\lambda^{\prime}}
\label{Ilambda'}
\end{equation}
Our approach is as follows. We will be interested in the amplitude of the
brane mode, $\psi_0$, towards the end of inflation. To this end, we will
treat the right hand sides of Eqs.~(\ref{psi_0}) and (\ref{psi_lambda}) as
perturbations. This approximation is certainly {\it not} valid at very
early times, when $P(\eta) \equiv k/a(\eta) \buildrel > \over {_{\sim}}
P_{LV}$: at those times the very notion of the brane mode does not make
sense. However, for $P_{LV} \gg H$ the relevant modes are in the adiabatic
regime at those times, mixing between positive- and negative-frequency
components of the field is negligible, and the field remains in its
adiabatic vacuum state. This point (mode generation in the adiabatic
regime) has been discussed in more detail in Ref.~\cite{mlvr}. At later
times, perturbation theory {\it is} justified by the fact that $U(y)$
peaks near the brane, where $\phi_\lambda (y)$ are suppressed; hence the
overlap integrals (\ref{Ilambda}) and (\ref{Ilambda'}) are small. In the
zeroth approximation, with overlaps neglected, equations for the brane
mode and bulk modes decouple and can be straighforwardly solved. At this
level, equation for the brane mode is
\[
\ddot{\psi}_0 - \frac{\ddot{a}}{a} \psi_0 + k^2 \psi_0 = 0
\]
It exactly coincides with the corresponding equation in
four-dimensional theory, so the zeroth order result for the spectrum is
exactly the same as in four dimensions. Namely, towards the end of
inflation, $\psi_0^{(0)} ({\bf k}, \eta)$ (the superscript here refers to
the zeroth order approximation) is a Gaussian field with the correlation
function
\[
\langle \psi_0^{(0)} ({\bf k}), \psi_0^{(0)} ({\bf k^\prime}) \rangle =
a^2 (\eta) \frac{2\pi^2}{k^3} {\cal P}^{(0)} (k) \delta^3 ({\bf k} - {\bf
k^\prime})
\]
with
\[
{\cal P}^{(0)} = \frac{H_k^2}{4\pi^2}
\]
and
\[
H_k = H (\eta_k)
\]
where $\eta_k$ is the time at which the mode of momentum $k$ crosses
out the horizon,
\[
H (\eta_k) = \frac{k}{a(\eta_k)}
\]
Thus, four-dimensional behavior of the field in the brane mode is
trivially obtained in the zeroth approximation.

The first non-trivial contribution to the field in the brane mode occurs
at the first order of perturbation theory, and is due to the overlap with
bulk modes in the right hand side of Eq.~(\ref{psi_0}). To evaluate this
contribution, we first have to solve Eq.~(\ref{psi_lambda}) at the zeroth
order.

At the zeroth order, Eq.~(\ref{psi_lambda}) becomes
\[
\ddot{\psi}_\lambda - \frac{\ddot{a}}{a} \psi_\lambda + \epsilon^2 k^2
\psi_\lambda + a^2 \lambda^2 \psi_\lambda = 0
\]
In the asymptotic past, the second and fourth terms in this equation
are negligible, the field is in the adiabatic regime, and we immediately
write its decomposition in creation and annihilation operators,
\begin{equation}
  \psi_\lambda^{(0)} = \frac{1}{\sqrt{2 \epsilon k}} (\psi_\lambda^{+}
  (\eta) A^+_{\lambda, k} + {\mbox h.c.})
\label{psi_lambda^0}
\end{equation}
where in the asymptotic past
\[
\psi_\lambda^+ = \mbox{e}^{i\epsilon k \eta} \; , \;\;\; \eta \to
-\infty
\]
As we will see in the next subsection, the main effect on the brane
mode comes from the bulk modes with
\begin{equation}
  \lambda \ll H
\label{assum2}
  \end{equation}
  where $H$ is, roughly speaking, the inflationary Hubble parameter. These
  modes get out from the adiabatic regime at the time $\eta_{\epsilon k}$
such that
\[
H (\eta_{\epsilon k}) \equiv H_{\epsilon k} \sim \frac{\epsilon
k}{a(\eta_{\epsilon k})}
\]
Let us call this moment of time ``$\epsilon$-horizon crossing''. One of
our key observations is that for small $\epsilon$, this moment occurs much
earlier than the horizon crossing by the brane mode. In terms of the
original field $\Phi $, the bulk modes  thus freeze out at
much larger amplitudes than the brane mode, and their effect on brane mode
is enhanced.

Immediately after $\epsilon$-horizon crossing, the bulk mode behaves as
\begin{equation}
   \psi_\lambda^+ (\eta) = a(\eta) \frac{H_{\epsilon k}}{\epsilon k} \; ,
   \;\;\;\; \eta \buildrel > \over {_{\sim}} \eta_{\epsilon k}
\label{inpsi}
\end{equation}
We will need this mode at the moment $\eta_k$ at which the brane mode
crosses out the horizon. For small $\epsilon$ this moment occurs much
later than $\epsilon$-horizon crossing, so some care must be taken at this
point. The easiest way to proceed is to make use of the original equation
(\ref{eqphi1}) for the field $\phi$. After $\epsilon$-horizon crossing,
the field is in the slow roll regime and the physical momentum is small,
so that the first and third terms in Eq.~(\ref{eqphi1}) are negligible.
The field $\phi(t)$ thus evolves as follows,
\[
\phi (t) \propto \mbox{exp} \left(-\int_{t_{\epsilon k}}^t~
dt^\prime \frac{\lambda^2}{3H(t^\prime)}\right)
\]
Making use of the initial condition (\ref{inpsi}), one obtains in terms
of $\psi_\lambda$
\[
  \psi_\lambda^+ (\eta) = a(\eta) \frac{H_{\epsilon k}}{\epsilon k}
  \mbox{exp} \left(-\int_{\eta_{\epsilon k}}^\eta~d\eta^\prime
\frac{\lambda^2}{3H(\eta^\prime)} a(\eta^\prime)\right) \; , \;\;\; \eta
\gg \eta_{\epsilon k}
  \]
Let us introduce the following ``mean value'' $\hat{H}$ of the
inflationary Hubble parameter in the interval $(\eta_{\epsilon k},
\eta_k)$
\begin{eqnarray}
\frac{1}{\hat{H}^2} &=& \frac{1}{|\log \epsilon|} \int_{\eta_{\epsilon
k}}^{\eta_k}~d\eta^\prime \frac{1}{H(\eta^\prime)} a(\eta^\prime)
\nonumber \\
&=& \frac{1}{|\log \epsilon|} \int_{a(\eta_{\epsilon
k})}^{a(\eta_k)}~\frac{da}{a} \frac{1}{H^2(a)}
\end{eqnarray}
For time-independent $H$ one has $\hat{H}=H$, while in general $\hat{H}$
is a non-trivial function of $\epsilon$ and $k$. In terms of this
parameter, the bulk mode behaves near $\eta_k$ as follows,
\begin{equation}
\psi_\lambda^+ (\eta) = a(\eta) \frac{H_{\epsilon k}}{\epsilon k}
\mbox{exp}\left(-\frac{\lambda^2}{3\hat{H}^2} |\log \epsilon|\right) \; ,
\;\;\; \eta \sim \eta_k
\label{end0}
\end{equation}
This completes the discussion of the zeroth approximation.

\subsection{Bulk contribution into brane mode}

We now wish to calculate the effect of the bulk modes on the brane mode,
in the lowest non-trivial order of perturbation theory. To this end, we
insert the zeroth order expression for the bulk modes,
Eq.~(\ref{psi_lambda^0}) into Eq.~(\ref{psi_0}) and obtain the equation
for the first correction to $\psi_0$,
\begin{equation}
\ddot{\psi}_0^{(1)} - \frac{\ddot{a}}{a} \psi_0^{(1)} + k^2 \psi_0^{(1)} =
- k^2 \int~d\lambda~\psi_\lambda^{(0)} I_\lambda
\label{firstordereq}
\end{equation}
We are interested in the solution to this equation with zero initial
condition at infinite past: all modes oscillate as $\eta \to -\infty$, and
mixing between the modes is negligible at that time. This solution is
\begin{equation}
\psi_0^{(1)} (\eta) = - k^2 \int~d\eta^\prime G(\eta, \eta^\prime)
\int~d\lambda~\psi_\lambda^{(0)} (\eta^{\prime}) I_\lambda
\label{Eq/Pg14/1:paper}
\end{equation}
where $G$ is the retarded Green's function of the operator in the left
hand side of Eq.~(\ref{firstordereq}). We will see that the effect of
mixing is most relevant at $\eta \sim \eta_k$, so we approximate
\begin{equation}
     a(\eta) = -\frac{1}{H_k \eta}
\label{aetak}
     \end{equation}
     and obtain
\[
     G(\eta, \eta^\prime) = \theta (\eta - \eta^\prime) F(\eta, \eta^\prime)
     \]
    where
    \[
    F(\eta, \eta^\prime) = \frac{1}{2ik} \mbox{e}^{ik(\eta -
    \eta^\prime)} \left(1 + \frac{i}{k\eta}\right) \left(1 -
\frac{i}{k\eta^\prime}\right) + \mbox{c.c.}
    \]
Our purpose is to calculate the amplitude of the brane mode after
the horizon crossing, $\eta \gg \eta_k$. We find in this region
\[
F(\eta, \eta^\prime) = -a(\eta) \frac{H_k}{k^2} \left(\cos
(k\eta^\prime) - \frac{\sin(k\eta^\prime)}{k \eta^\prime} \right) \; ,
\;\;\; \eta \gg \eta_k
\]
The integrand in Eq.~(\ref{Eq/Pg14/1:paper}) rapidly oscillates at
$\eta^\prime \ll \eta_k$ and decays at $\eta^\prime \gg \eta_k$. Thus, the
integral is saturated at $\eta^\prime \sim \eta_k$. The bulk modes at that
time have the form (\ref{end0}), where one can again use the relation
(\ref{aetak}). Combining all factors and performing the integration over
$\eta^\prime$ we obtain at $\eta \gg \eta_k$
\begin{equation}
\psi_0^{(1)} (\eta) = - a(\eta) \frac{1}{\sqrt{2}} \frac{H_{\epsilon
k}}{\epsilon^{\frac{3}{2}} k^{\frac{3}{2}}} \int~d\lambda~ I_\lambda
\mbox{e}^{-\frac{\lambda^2}{3\hat{H}^2} |\log \epsilon|} A^+_\lambda +
\mbox{h.c.}
\label{psi_0^1}
\end{equation}
This is again a Gaussian field whose spectrum is
\[
{\cal P}^{(1)} = \frac{H^2_{\epsilon k}}{4\pi ^2\epsilon^3}
\int~d\lambda~ |I_\lambda|^2 \mbox{e}^{-\frac{2\lambda^2}{3\hat{H}^2} |\log
\epsilon|}
\]
Because of the relation (\ref{Ineq}), the integral here is convergent
at small $\lambda$, and therefore it is saturated at
\[
\lambda \sim \lambda_c = \frac{\hat{H}}{|\log \epsilon|^{1/2}}
\]
This justifies our assumption (\ref{assum2}). We obtain finally
\begin{equation}
{\cal P}^{(1)} = \mbox{const} \cdot \frac{H^2_{\epsilon k}}{\epsilon^3}
\lambda_c |I_{\lambda_c}|^2
\label{P2}
\end{equation}
where the constant is of order 1. This is our final result.

Several remarks are in order. First, the Gaussian field (\ref{psi_0^1}) is
independent of the field $\psi_0^{(0)}$ considered in the previous
subsection, as the latter contains creation and annihilation operators in
the incoming brane mode. Thus, towards the end of inflation, perturbations
on the brane are the sum of two independent Gaussian fields with spectra
${\cal P}^{(0)}$ and  ${\cal P}^{(1)}$. Second, because of the relation
(\ref{Ineq}), the contribution (\ref{P2}) indeed tends to zero as $P_{LV}
\to \infty$. However, the suppression at small $H/P_{LV}$ may be quite
weak, depending on the form of the warp factors. As an example, for the
choice (\ref{ab}) the suppression factor is (see Eqs.~(\ref{lIl}) and
(\ref{ybP}))
\[
\lambda_c I^2_{\lambda_c} \sim \left(\frac{\hat{H}}{P_{LV}}
\right)^{2\nu -2}
\]
(up to $\log \epsilon$), while the only restriction on the parameter
$\nu$ is $\nu >1$. Third, the bulk-induced contribution to the spectrum is
enhanced by $\epsilon^{-3}$; we have already discussed the origin of this
enhancement in previous subsection. Fourth, the overall magnitude of
${\cal P}^{(1)}$ is determined by $H_{\epsilon k}$, the value of the
Hubble parameter at $\epsilon$-horizon crossing, which occurs earlier than
the usual horizon crossing by the brane mode, and also by $\hat{H}$, which
is a certain average of the Hubble parameter over rather long time. Thus,
the two contributions to the spectrum, ${\cal P}^{(0)}$ and ${\cal
P}^{(1)}$ generically have different tilts. Finally, it is straightforward
 to check that both the back reaction of the brane mode on bulk modes
(first integral in the right hand side of Eq.~(\ref{psi_lambda})) and
mixing between bulk modes (second integral) indeed give small effects,
even if ${\cal P}^{(1)} > {\cal P}^{(0)}$, so our perturbative treatment
is justified.

\section*{ACKNOWLEDGMENTS} The authors are indebted to D.~Krotov,
D.~Levkov and D.~Semikoz for helpful discussions. This work
is supported in part by RFBR grant 05-02-17363-a, by the Grants of the
President of Russian Federation NS-2184.2003.2, MK-3507.2004.2, by INTAS
grant YSF 04-83-3015, and by the grant of the Russian Science Support
Foundation.

\end{document}